%% file: main.tex
\documentclass[sigconf]{acmart}

\AtBeginDocument{%
  }

\copyrightyear{2026}
\acmYear{2026}
\setcopyright{cc}
\setcctype{by}
\acmConference[UIST '26]{The 39th Annual ACM Symposium on User Interface Software and Technology}{November 02--05, 2026}{Detroit, MI, USA}
\acmBooktitle{The 39th Annual ACM Symposium on User Interface Software and Technology (UIST '26), November 02--05, 2026, Detroit, MI, USA}
\acmDOI{10.1145/3830398.3830621}
\acmISBN{979-8-4007-2856-3/2026/11}

\usepackage{tabularx} 
\usepackage{booktabs}
\usepackage{multirow} 
\usepackage{float} 
\usepackage{xcolor}
\usepackage{graphicx}

\newcommand{\sysname}{\textsc{Maru}}
\newcommand{\username}{Max} 
\newcommand{\bryan}[1]{}
\newcommand{\figlabel}[1]{%
\begingroup
\setlength{\fboxsep}{1pt}%
\fcolorbox{black}{black}{\textcolor{white}{\scriptsize #1}}%
\endgroup
}

\usepackage{color}
\definecolor{Silver}{rgb}{0.752,0.752,0.752}
\usepackage{float} 

\newif\ifhighlight\highlightfalse

\definecolor{revised}{HTML}{030362} 

\newcommand{\revised}[1]{%
  \ifhighlight
    \textcolor{revised}{#1}%
  \else
    #1%
  \fi
}

\definecolor{Issue1}{HTML}{148f6a} 
\definecolor{Issue2}{HTML}{4860cc}
\definecolor{Issue3}{HTML}{247e1d}
\definecolor{Issue4}{HTML}{d52b2c}
\definecolor{Issue5}{HTML}{9367bc}
\definecolor{Issue6}{HTML}{4b3c88}
\definecolor{Issue7}{HTML}{ad4d33}
\definecolor{Issue8}{HTML}{a4008b}
\definecolor{Issue9}{HTML}{bf741f}
\definecolor{Issue10}{HTML}{0a8970}
\definecolor{MISC}{HTML}{868686}

\newcommand{\pointmark}[3]{%
  \ifhighlight
    \colorbox{#1}{\textcolor{white}{\footnotesize\bfseries \##2}}\textcolor{#1}{#3}%
  \else
    #3%
  \fi
}

\newcommand{\pointone}[1]{\pointmark{Issue1}{1}{#1}}
\newcommand{\pointtwo}[1]{\pointmark{Issue2}{2}{#1}}
\newcommand{\pointthree}[1]{\pointmark{Issue3}{3}{#1}}
\newcommand{\pointfour}[1]{\pointmark{Issue4}{4}{#1}}
\newcommand{\pointfive}[1]{\pointmark{Issue5}{5}{#1}}
\newcommand{\pointsix}[1]{\pointmark{Issue6}{6}{#1}}
\newcommand{\pointseven}[1]{\pointmark{Issue7}{7}{#1}}
\newcommand{\pointeight}[1]{\pointmark{Issue8}{8}{#1}}

\begin{document}

\title[Maru: Information Architecture for Generative UI]{Maru: Information Architecture as a Shared Language for Generating Aligned and Persistent User Interfaces}


\author{Eunhye Kim}
\orcid{0009-0004-1460-8532}
\affiliation{%
  \institution{School of Computing, KAIST}
  \city{Daejeon}
  \country{Republic of Korea}
}
\email{gracekim027@kaist.ac.kr}

\author{DaEun Choi}
\orcid{0000-0002-4843-0486}
\affiliation{%
  \institution{School of Computing, KAIST}
  \city{Daejeon}
  \country{Republic of Korea}
}
\email{daeun.choi@kaist.ac.kr}

\author{Bryan Min}
\orcid{0009-0003-0657-4398}
\affiliation{%
  \institution{University of California San Diego}
  \city{La Jolla}
  \state{California}
  \country{USA}
}
\email{bdmin@ucsd.edu}

\author{Hyunjung Yi}
\orcid{0009-0000-6617-3105}
\affiliation{%
  \institution{Korea University}
  \city{Seoul}
  \country{Republic of Korea}
}
\email{ruby3672@korea.ac.kr}

\author{Yue Jiang}
\orcid{0000-0003-0022-6512}
\affiliation{%
  \institution{University of Utah}
  \city{Salt Lake City}
  \state{Utah}
  \country{USA}
}
\email{yue.jiang@utah.edu}

\author{Juho Kim}
\orcid{0000-0001-6348-4127}
\affiliation{%
  \institution{School of Computing, KAIST}
  \city{Daejeon}
  \country{Republic of Korea}
}
\affiliation{%
  \institution{SkillBench}
  \city{Santa Barbara}
  \state{California}
  \country{USA}
}
\email{juhokim@kaist.ac.kr}

\renewcommand{\shortauthors}{Kim et al.}

\begin{abstract}
\input{sections/00_abstract}

\end{abstract}


\begin{teaserfigure}
\centering
    \includegraphics[width=\linewidth]{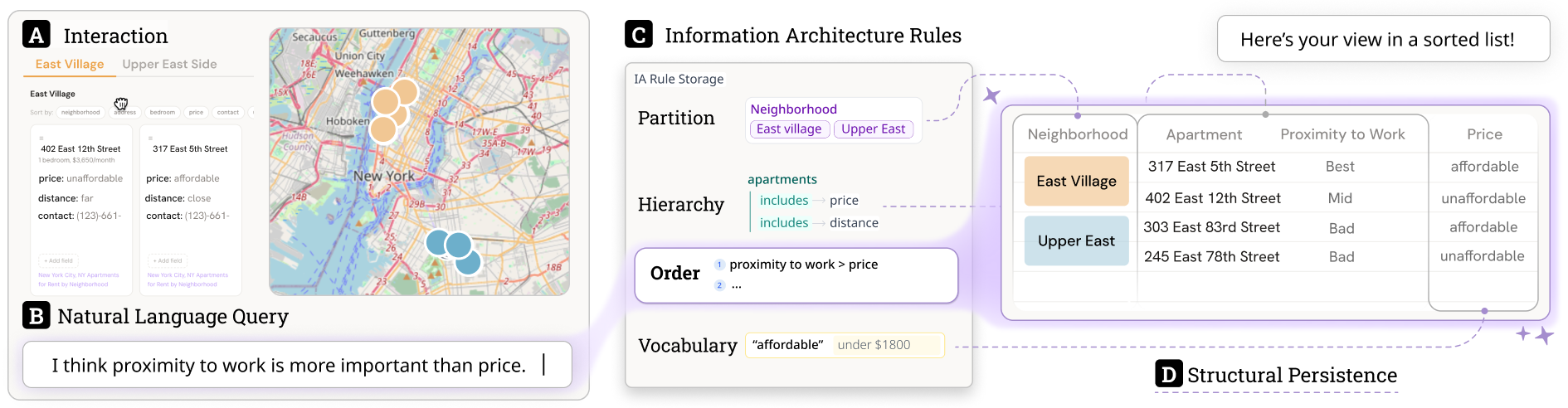}
    \caption{\sysname{} carries forward what users establish as IA rules across generations. A user browses apartments in a map+tabs view and states \textit{"I think proximity to work is more important than price,"} (B) adding a new Order rule to their accumulated IA state. \pointeight{The same rule could be created by sorting the list directly in the generated UI (A)}. \sysname{} surfaces the full IA state (C), with Partition, Hierarchy, Order, and Vocabulary rules captured across prior interactions and the new Order rule highlighted. The next generation produces a sorted table reflecting structural persistence (D). \textit{What changed:} the layout type and item ordering. \textit{What persisted:} the neighborhood partition, hierarchy fields, and affordability vocabulary, carried forward without re-specification.}
    \Description{A four-panel diagram illustrating how Maru persists Information Architecture (IA) rules across UI generations. (A) Interaction panel: A map and tabs view showing two apartment listings in East Village and Upper East Side, with attributes including price, distance, and contact. A cursor icon indicates drag interaction. (B) Natural Language Query panel: A text input field containing the query 'I think proximity to work is more important than price.' (C) Information Architecture Rules panel: The IA Rule Storage displays four rule types — Partition (Neighborhood: East village, Upper East), Hierarchy (apartments includes price, includes distance), Order (proximity to work > price, highlighted in purple), and Vocab ('affordable' defined as under \$1800). Dashed arrows connect rules to the output. (D) Structural Persistence panel: A sorted table with columns Neighborhood, Apartment, Proximity to Work, and Price. Rows are grouped by East Village and Upper East neighborhood partitions. Apartments are ranked by proximity to work, with affordability labels applied using the vocabulary rule.
}
    \label{fig:teaser}
\end{teaserfigure}


\maketitle

\input{sections/01_introduction}

\input{sections/02_relatedworks}

\input{sections/03_framework}

\input{sections/04_system}
\input{sections/05_evaluation}

\input{sections/06_results}

\input{sections/07_discussion}
\begin{acks}
This work was supported by Institute of Information \& Communications Technology Planning \& Evaluation (IITP) grant funded by the Korea government (MSIT) (No.2021-0-01347, Video Interaction Technologies Using Object-Oriented Video Modeling). This work was also supported by the National Research Foundation of Korea (NRF) grant funded by the Korea government (MSIT) (No.RS-2024-00406715).
We thank all our study participants for their time and valuable insights. The first author sincerely thanks all team members for their unwavering support and dedication. 
\end{acks}

\bibliographystyle{ACM-Reference-Format}
\bibliography{references_clean}

\appendix
\input{sections/99_appendix}

\end{document}

%% file: sections/00_abstract.tex


Generative user interfaces (GenUIs) promise on-demand components tailored to users’ needs. As users iterate on information tasks, they construct personal structures over information they encounter---how items are grouped, what gets prioritized, and what terms mean in their context. Yet, current systems leave these structural decisions to the model at each generation, ignoring the structural logic users have established. Without a persistent representational structure shared between user and system, GenUIs have no basis to remain aligned with what users have established.
We draw on Information Architecture (IA), a design practice for organizing and structuring information, as a shared language to bridge user-constructed structure and system generation.
We present a framework identifying four IA elements---partition, hierarchy, order, and vocabulary---and characterize how each maps to concrete UI generation decisions. We instantiate this framework in \sysname{}, a conversational system that captures user prompts and interactions as IA preferences, persisting as rules both user and system draw on across generations.
\pointseven{A user study revealed that IA persistence kept generated UIs aligned as sessions progressed, while alignment without it degraded}, with diverse patterns emerging across users and contexts, pointing to the value of IA persistence in aligning GenUI to individual needs.

%% file: sections/01_introduction.tex
\section{Introduction}

Advances in large language models (LLMs) are making the vision of  generative UI (GenUI) a reality---systems that can produce structured, personalized user interfaces on demand. Given a natural language query, these systems extract relevant information and render it into widgets, comparison tables, or interactive layouts tailored to the user's context~\cite{cao2025generative, leviathan2025generative, google2025disco, anthropic2024artifacts}.
This capability is shaping GenUI as an exciting opportunity towards achieving personal software across commercial tools, websites, and applications \cite{kay1977personal, litt2025malleable}. 

To make GenUI usable for real-world websites and commercial software, communities in industry, AI, and HCI have been exploring how to improve the quality of generated UIs~\cite{wu2024uicoder, jiang2025iluvui}, breadth of types of UIs that LLMs can generate \cite{anthropic2024artifacts, leviathan2025generative, chen2025generative}, and customizability of the generated UIs \cite{cao2025generative, google2025a2ui}.  Yet these advances say little about which UI elements should change or stay between prompts, and how much importance each element carries for the user throughout their task.

Although GenUI's ability to modify and regenerate nearly any aspect of an interface makes it highly customizable, this same flexibility introduces risks in modifying aspects of the UI that the user has come to rely on. For example, suppose a user receives a generated table listing available apartments in New York and finds the column-and-row structure personally valuable for comparing options. The system has no way of recognizing this value, and cannot guarantee that the next generated UI will retain it. As a result, users must re-articulate their UI preferences with each generation. Thus, to build GenUI systems that stay aligned with users over time, we must design systems that capture and preserve UI elements that the user values across generations.

In this paper, we aim to identify and provide this persistence in GenUI by drawing from the practice of \textbf{Information Architecture} (IA). IA is a UX design practice that operationalizes the mapping of user needs to concrete UI content and layouts~\cite{rosenfeld2002information, brown2010eight, guizani2022decade}. An IA design process for an apartment-finding interface might involve understanding the user's key criteria, prioritizing them to surface the most valued ones, grouping related content into navigable sections, and structuring the information into a list and map for users to explore.
We propose that IA offers a promising abstraction for this problem: specific enough to map directly onto concrete UI decisions, yet general enough to capture the structural logic users build across a task, and crucially, \textit{explicit} enough to be \textit{inspectable} and \textit{editable} by users. 
Our goal is to investigate whether integrating IA into the generation pipeline can enable the system to carry forward what users have established, producing UIs that remain aligned across generations.

We conducted a systematic codebook analysis of 42 papers spanning sensemaking theory and interactive systems, from which we derived four IA elements: \textbf{partition}, \textbf{hierarchy}, \textbf{order}, and \textbf{vocabulary}. We then characterized how each maps into concrete generation decisions, and then instantiated this framework in \sysname{}, a probe that captures IA rules from natural language prompts and direct interactions with generated outputs, and applies them as a persistent layer across generations.
Finally, we conducted a user study (N=12), observing when  persistence helped or hindered users, how users utilized the IA rules, and how they helped them personalize the UI.

\revised{Our user study found that \sysname{} maintained approval rates across sessions (74\% to 61\%) while the baseline collapsed from 71\% to 33\%, that 97\% of user-created rules were built through normal interaction and query behavior rather than deliberate schema editing, and that \sysname{} produced distinct layout types where baseline participants converged on similar ones.} However, IA persistence degraded alignment during sub-task transitions and when rules over-accumulated beyond relevance.
\pointseven{Overall, our results suggest that integrating IA rules into the UI generation pipeline helps users shape the interface to their needs, and that persistence needs boundaries to stay useful.} \revised{We open-source \sysname{}, including the IA layer, in hopes that it can support others in building GenUI systems that carry user-constructed structure forward across generations.\footnote{\url{https://github.com/kixlab/Maru}}}



\bryan{Then bullet point list of contributions -- make it concise}
In summary, we contribute:
\begin{itemize}
    \item A framework of four IA elements, mapping user behavior in sensemaking to UI generation decisions.
    
    \item \sysname{}, a probe instantiating this framework as an explicit, user-controllable IA layer for generative UI.
    
    \item Empirical findings on how IA persistence shapes UI alignment, user engagement, and personalization.
\end{itemize}

%% file: sections/02_relatedworks.tex
\section{Related Work}

\subsection{Achieving Personal Software with Generative UI}

Traditional GenUI approaches primarily involved breaking down key UI components into modular elements assembled automatically. Model-Based User Interface Design (MBUID) developed high-level models mapping to UI specifications, describing task stages~\cite{kasik1982user, pinheiro2000user, sukaviriya1994model} or the data, functionality, and view~\cite{benson2013cascadingtree}, enabling the system to translate high-level user needs into UI components \cite{gajos2004supple, nichols2006uniform, nichols2006huddle, nichols2002generating, nichols2004improving}. \pointone{These models were authored before deployment and, by design, remained hidden from end-users.} Other approaches generated UIs from interaction feedback rather than models~\cite{vaithilingam2019bespoke}. Although these approaches could generate rich, interactive UIs, a common barrier was domain \revised{specificity}: these systems were limited to dialog UIs \cite{gajos2004supple}, command-line GUIs \cite{vaithilingam2019bespoke}, or predefined templates \cite{nichols2002generating, nichols2004improving, luo1993management}.

LLMs removed this barrier by enabling UI generation across domains---with chatbots generating UI components and widgets \cite{anthropic2024artifacts, vaithilingam2024dynavis} and SVGs \cite{anthropic2026claudevisuals}, and full front-end prototypes \cite{google2025disco, v0vercel2023, lovable2026}. This capability is also advancing software customization and end-user programming through UI mashups \cite{zhang2018fusion, lin2009endvegemite}, shareable and reusable UI components \cite{klokmose2015webstrates, gronbaek2021mirrorblender, gobert2023lorgnette}, and personalized UI styles \cite{kim2022stylette}.

However, a key challenge for LLM-powered GenUI systems is keeping generated UIs aligned to the user's needs across prompt iterations as the user's task progresses. A general approach the HCI community has pursued involves generating intermediate outputs before the final UI to enrich the LLM's context. Recent work proposes generating task-driven data models~\cite{cao2025generative,xu2025duetui}, while other work infers user needs ``in-the-moment'' to produce more contextually relevant UIs at each generation stage \cite{lam2025just,shaikh2025creating}.

However, these approaches still delegate to the LLM the decision of which UI elements to generate from context. Our work instead makes this mapping explicit. We provide a structured encoding that determines which elements to persist and which to change, giving end-users direct visibility and control over how their needs map to UI elements. \pointone{In doing so, our work joins a broader tradition of declarative specifications for interfaces~\cite{satyanarayan2016vega, beaudouin2026belidor}, which differ from one another in the semantics that they encode---in our case, how a user organizes information.}

\subsection{Designing for Persistence in AI Systems}





Persistence has long been recognized as essential for continuity in information-intensive tasks. Users actively construct internal structures---schemas, criteria, priorities, and relational maps---to make sense of what they encounter~\cite{pirolli1995information, klein2006making, pirolli2005sensemaking}, with significant effort directed toward finding the right representational schema rather than consuming content~\cite{russell1993cost, kittur2013costs, piorkowski2013whats}. Early work addressed this through structured representations: sensemaking systems demonstrated that externalizing these structures reduces reconstruction cost~\cite{liu2024selenite, chang2020mesh, chang2019searchlens, suh2023sensecape}, and that exploratory search advances as users build increasingly rich organizational structures~\cite{marchionini2006exploratory, rieh2016towards}. However, externalizing structure remains effortful even with lightweight tools~\cite{liu2022wigglite, chang2021tab, hahn2018bento, nguyen2016sensemap, roy2021note, kuznetsov2024tasks}, and the structures captured remain session-specific, and cannot transfer beyond the tool's own immediate affordances~\cite{yun2025generative}.

LLM-era work has approached persistence more flexibly, inferring preferences~\cite{shaikh2025creating, mei2025interquest}, conversational memories~\cite{yen2024memolet}, and intent specifications~\cite{zhao2025knoll, vaithilingam2025semantic} from interaction data. Across these approaches, the underlying motivation is consistent: without carry-forward, users must re-establish context, and alignment degrades over time. Systems have shown that making accumulated context visible and editable gives users greater confidence and control~\cite{yen2024memolet, vaithilingam2025semantic}, \pointthree{and that surfacing what the system has inferred or would infer differently helps users refine their intent~\cite{zamfirescu2025pail, gebreegziabher2025mocha}.}

What remains underexplored is persistence at the structural level of generated interfaces, where LLM-era approaches carry context fluidly but were not designed to preserve the structural preferences that shape what a useful interface looks like for a given user. Our work addresses this gap directly for GenUI.

\subsection{Information Architecture for UI Persistence}





Information architecture (IA) is a UX design practice concerned with organizing, labeling, and structuring information for effective navigation and sensemaking~\cite{rosenfeld2002information, brown2010eight, guizani2022decade}. In practice, IA work begins with understanding user needs and making structural decisions: how content is grouped, what hierarchies govern navigation, how items are sequenced, and what terminology is used~\cite{rosenfeld2002information}. These decisions are externalized through sitemaps, taxonomies, and card sorting exercises that resolve diverse organizational schemes into a single canonical structure~\cite{katsanos2008autocardsorter, brunetti2013design}. The result is a set of structural rules established before any UI is rendered, which acts as a blueprint shaping how users explore and understand information~\cite{sarrafzadeh2016knowledge, brunetti2013design, schaik2015automated}. When interface architecture conflicts with a user's own logic it creates information anxiety; when it aligns, it acts as a cognitive scaffold~\cite{ohlin2012role, sen2006tagging}.





In traditional practice, these decisions are author-owned and static---settled at design time and applied uniformly to users~\cite{rosenfeld2002information, brown2010eight, guizani2022decade}. Recent work challenges this: research on malleable interfaces show that end-users produce diverse and personally meaningful arrangements when given control over content and layout~\cite{min2025malleable, min2025meridian, alves2024citizen}, and adjacent systems have proposed structured layers that sit alongside generation~\cite{cao2025generative}. Yet in current GenUI systems, the mapping from user context to UI decisions is delegated entirely to the LLM, which neither exposes nor persists the structural logic it applies. Thus, the user has limited ability to inspect, correct, or build on it.

\pointone{We propose that IA offers a promising abstraction for this problem. Unlike specifications written for a system, IA describes how information is organized for a person to make sense of it, which is a concern users already have about their own information.} By making structural mapping explicit, persistent, and user-editable, IA can serve as an enabling layer through which the system captures structural rules derived from user behavior, and the user can inspect and modify those same rules to further shape the interface~\cite{spagnolo2010beyond, ohlin2012role, hinton2009machineries}.

%% file: sections/03_framework.tex
\section{A Framework of IA Elements for GenUI}


The goal of this framework is to guide the use of IA elements in mapping user needs and activities to concrete UI patterns. We built this framework through three steps: (1) collecting literature and tools, (2) identifying the user needs and resulting UI patterns supported by each system, and (3) synthesizing common elements that recur across systems as mappings between user behavior and UI structure. 


In this paper, we ground our framework in sensemaking literature. Sensemaking underlies a wide range of everyday information tasks such as conversing with LLMs or searching, and the research around it has theorized both sides of this mapping---how users actively build structure organizing information and how interfaces support that~\cite{russell1993cost,pirolli1995information,klein2006making}. This makes sensemaking literature a principled and broadly applicable source for deriving IA elements that can serve as a shared language between user behavior and system generation. \pointtwo{Applying the same approach to other domains such as writing~\cite{lam2025just, chen2025genui}, data analysis, or learning~\cite{chen2025generative} would likely yield a different set of elements. Generative systems for other domains surface structures organized around how behavior decomposes or how variations branch, rather than how information is organized~\cite{aveni2025generative, angert2023spellburst}. We therefore scope our framework to sensemaking, where user-constructed structure is central, and view the four elements as a starting vocabulary for identifying what persists in other domains.}


\subsection{Derivation Method}
We conducted a thematic analysis of existing systems, theories, and empirical studies. We included papers that describe systems or interaction techniques supporting the organization or navigation of information during information tasks, or provide theoretical or empirical accounts of user behavior during such tasks. 

Our analysis proceeded in three phases. Two coders independently analyzed an initial set of 12 seed papers spanning seminal sensemaking theory and systems, developing an initial codebook on the user needs, UI elements, and interactions supported through discussion. They then coded 8 additional papers identified through snowball sampling, refining and stabilizing the codebook until no new codes emerged. A single coder then applied the finalized codebook to the remaining 22 papers, reaching a total of 42 papers. For each IA element, a full mapping of the coded systems, their supported interactions, and corresponding UI patterns is provided in Appendix~\ref{tab:literature-codebook}, and the final set of papers is provided in Appendix~\ref{tab:literature-set}.

\subsection{Defining and Characterizing the Elements}

Our analysis yielded four IA elements---partition,  hierarchy, order, and vocabulary---along with a set of system support mechanisms and visual/interaction patterns.
The four IA elements are divided into two categories: structural elements (partition and hierarchy), which describe how users organize information, and semantic elements (order and vocabulary), which capture the meaning and criteria users bring to that structure. 
 
To illustrate our framework, we use an example of an apartment search task, in which a user browses multiple listings, reads reviews, and compares options across several sessions. 

\begin{figure*}[h]
    \centering
    \includegraphics[width=1\linewidth]{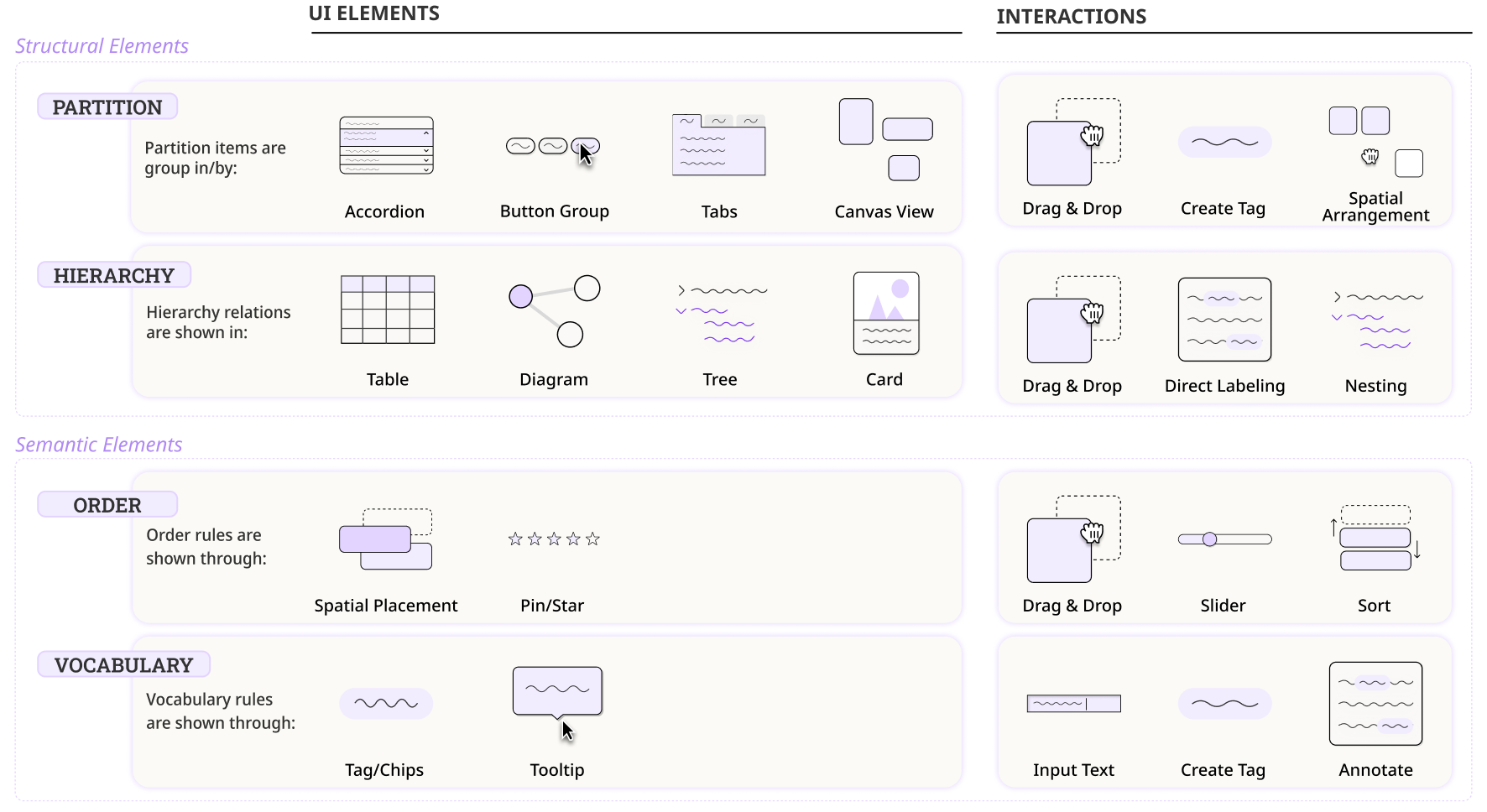}
    \caption{Our framework of IA elements, with representative UI components and construction interactions for each element. \pointeight{Each element can be both displayed and constructed through the UI itself. A partition rule, for instance, can be shown as a set of tabs grouping related items, and edited by dragging an item from one tab to another, which moves it between groups.}}
    \Description{A framework diagram organized into two main columns — UI Elements and Interactions — and four rows corresponding to the four IA elements. The top two rows are labeled Structural Elements, and the bottom two are labeled Semantic Elements. Partition row: UI elements include Accordion (icon of a collapsible list), Button Group (icon of pill-shaped toggle buttons), Tabs (icon of a tabbed panel), and Canvas View (icon of free-floating rectangles). Interactions include Drag & Drop (cursor dragging a card), Create Tag (cursor near a tag shape), and Spatial Arrangement (cursor repositioning rectangles). Hierarchy row: UI elements include Table (grid icon), Diagram (node-link graph with one parent and two child nodes), Tree (indented text list), and Card (image card icon). Interactions include Drag & Drop, Direct Labeling (text label applied to a component), and Nesting (indented component within another). Order row: UI elements include Spatial Placement (two rectangles stacked with the top one emphasized) and Pin/Star (five outlined star icons). Interactions include Drag & Drop, Slider (a horizontal slider with a filled handle), and Sort (list with a downward arrow indicating reordering). Vocabulary row: UI elements include Tag/Chips (a small wavy label chip) and Tooltip (a popup bubble with a cursor). Interactions include Input Text (a text field with a wavy line), Create Tag, and Annotate (a bordered text box).
}
    \label{tab:element-ui-mapping}
\end{figure*}

\subsubsection{Partition}
Partition is an organizational structure in which items are grouped together because they are similar or related, without any item assuming a special role relative to another. Items within a partition are \textit{lateral peers}---they share a boundary but not a hierarchy.
In our apartment search example, a user might group listings into neighborhoods (\textit{Downtown, Midtown, Uptown}) or divide their criteria into \textit{must-haves} and \textit{nice-to-haves}.

For example, in ForSense~\cite{rachatasumrit2021forsense}, users dragged information clips into color-coded spatial clusters on a canvas, regrouping them as their mental model of a topic evolved. Each cluster is a peer of the others with no group subordinating another. In SearchLens~\cite{chang2019searchlens}, users \revised{tagged} keywords into Lenses rendered as a button group, keeping each facet distinct and independently adjustable. Together these systems show that \revised{partition groups create boundaries that are explicit} without implying role relationships between them.

Beyond these, systems supported partition construction through drag-and-drop into groups~\cite{kuznetsov2022fuse, rachatasumrit2021forsense}, tag or label assignment~\cite{chang2019searchlens}, and spatial arrangement on a canvas~\cite{suh2023sensecape}. These were rendered through spatial clusters~\cite{rachatasumrit2021forsense, suh2023sensecape}, tab groups and card clusters~\cite{kuznetsov2022fuse, chang2021tabs}, color coding as a boundary signal~\cite{rachatasumrit2021forsense, chang2023citesee}, and faceted result sections~\cite{chang2019searchlens, fogarty2008cueflik}.

\subsubsection{Hierarchy}
Hierarchy is an organizational structure in which items bear differentiated roles relative to one another, encoding a \textit{semantic role relationship}---position signals \revised{how an item stands relative to its neighbors}. A user might place evidence clips about safety, commute time, and price under Apartment A, treating each clip as \textit{evidence} that characterizes the \textit{option}. Hierarchy applies to information items, partitions, and other hierarchies.

Tabs.do~\cite{chang2021tabs} illustrates hierarchy as decomposition: users nested task bundles under parent bundles through drag-and-drop, building persistent tree structures. The role relationship here is one of containment, where each level qualifies and contextualizes the one below it. Unakite~\cite{liu2019unakite} shows a different form of hierarchy, where users organized programming options against evaluation criteria in a comparison table, with evidence clips nested under their respective criteria and options. Here the role relationship is one of evidence: position signals what supports what. 

Beyond these, systems supported hierarchy construction through drag-and-drop into containers~\cite{liu2019unakite, kuznetsov2022fuse}, context menus for nesting~\cite{chang2021tabs}, and tag creation~\cite{yen2024memolet}. These were rendered through tables~\cite{liu2019unakite, chang2020mesh}, accordions and toggles~\cite{chang2021tabs, kang2023synergi}, nested cards~\cite{kuznetsov2022fuse, ward2021orgbox}, and node-link diagrams~\cite{yen2024memolet, palani2022interweave, suh2023sensecape}.

\subsubsection{Order}
Order is the assignment of \textit{sequence} or \textit{weight} to items, partitions, or hierarchies based on the user's subjective sense of importance, priority, or relevance. Unlike hierarchy and partition, order describes a \textit{property} of elements within a structure. In our apartment seeking example, a user might weight \textit{safety} as three times more important than \textit{commute time}, or rank Apartment A as their current top choice. 


Mesh~\cite{chang2020mesh} illustrates order as discrete relative weighting: users assigned importance multipliers to criteria and dragged rows to reflect their intuitive ranking, with options reordered by aggregate score so the most preferred rose to the top. The user's priorities are directly encoded as structural properties of the output. CueFlik~\cite{fogarty2008cueflik} shows a different form of order: users manipulated the relative importance of each active rule through per-rule weight sliders, with the final ranking of results continuously shifting to reflect the combined weighting. 

Beyond these, systems encoded order through drag-and-drop interactions~\cite{chang2020mesh, liu2019unakite}, priority selectors and pin or star actions~\cite{chang2021tabs, hahn2018bento}, and weight sliders~\cite{fogarty2008cueflik}. These were rendered through spatial placement reflecting importance~\cite{chang2020mesh, chang2021tabs}, weight badges and emphasis on high-priority items~\cite{chang2019searchlens, ruotsalo2014intentradar}, and collapsing or filtering of low-priority items~\cite{liu2022wigglite, chang2020mesh}.

\subsubsection{Vocabulary}
Vocabulary is the set of personal terms a user establishes with task-specific or session-specific meaning, functioning as a \textit{controlled personal lexicon}: once defined, terms serve as persistent semantic anchors for organizing, labeling, and evaluating information consistently across the task. A user might define \textit{affordable} as any listing under \$1,800/month---an evaluation criterion that applies consistently across all listings. 


CheatSheet~\cite{vermette2015cheatsheet} illustrates vocabulary well: users created and assigned task-specific tags through direct text input, building a personal navigational vocabulary that persisted across their note library. These terms were rendered as tag chips and surfaced as tooltips on hover, keeping the user's own terminology visible at the point of use.

Beyond this, systems supported vocabulary construction through inline text input~\cite{liu2019unakite, liu2022crystalline}, tag creation~\cite{liu2022wigglite, ros2024textdata}, and term extraction from annotated content~\cite{subramonyam2020texsketch, suh2023sensecape}. These were rendered through persistent label components using the user's own terminology~\cite{liu2019unakite, chang2020mesh}, tag chips and filter labels~\cite{liu2024selenite, liu2022wigglite}, and tooltips and inline highlights~\cite{chang2023citesee, zheng2024disciplink}.


\subsection{Persistence of IA Elements}
Because IA elements capture the structural and semantic preferences users build over the course of a task, they are persistent by nature. Yet our analysis found that most systems treat user-constructed structure as ephemeral, discarding it when a query ends or a session closes~\cite{liu2024selenite,suh2023sensecape,chang2020mesh}. Several systems documented this kind of persistence as a desired behavior that went unsupported~\cite{chang2019searchlens,fogarty2008cueflik,dontcheva2006summarizing}, while systems that did support it showed measurably lower reconstruction costs~\cite{chang2021tabs,chang2019searchlens}. We therefore define \textbf{\textit{persistence}} as the property of IA elements that carry forward across generations within a session and should be retrieved and reapplied to new generations, with the user being able to explicitly add or drop them.

%% file: sections/04_system.tex
\section{The \sysname{} Probe}
\revised{We instantiate the framework in \sysname{}, a chat-based GenUI system.}
\pointtwo{The IA layer is implemented as a set of skill specifications independent of \sysname{}'s UI components, so other GenUI systems can consume the rule schema directly without adopting our rendering pipeline.} \revised{Both the IA layer and the \sysname{} implementation are available at \url{https://github.com/kixlab/Maru}.}

\subsection{Envisioned User Scenario}
We introduce \sysname{} with an envisioned user scenario.

\begin{figure*}[h]
    \centering
    \includegraphics[width=\linewidth]{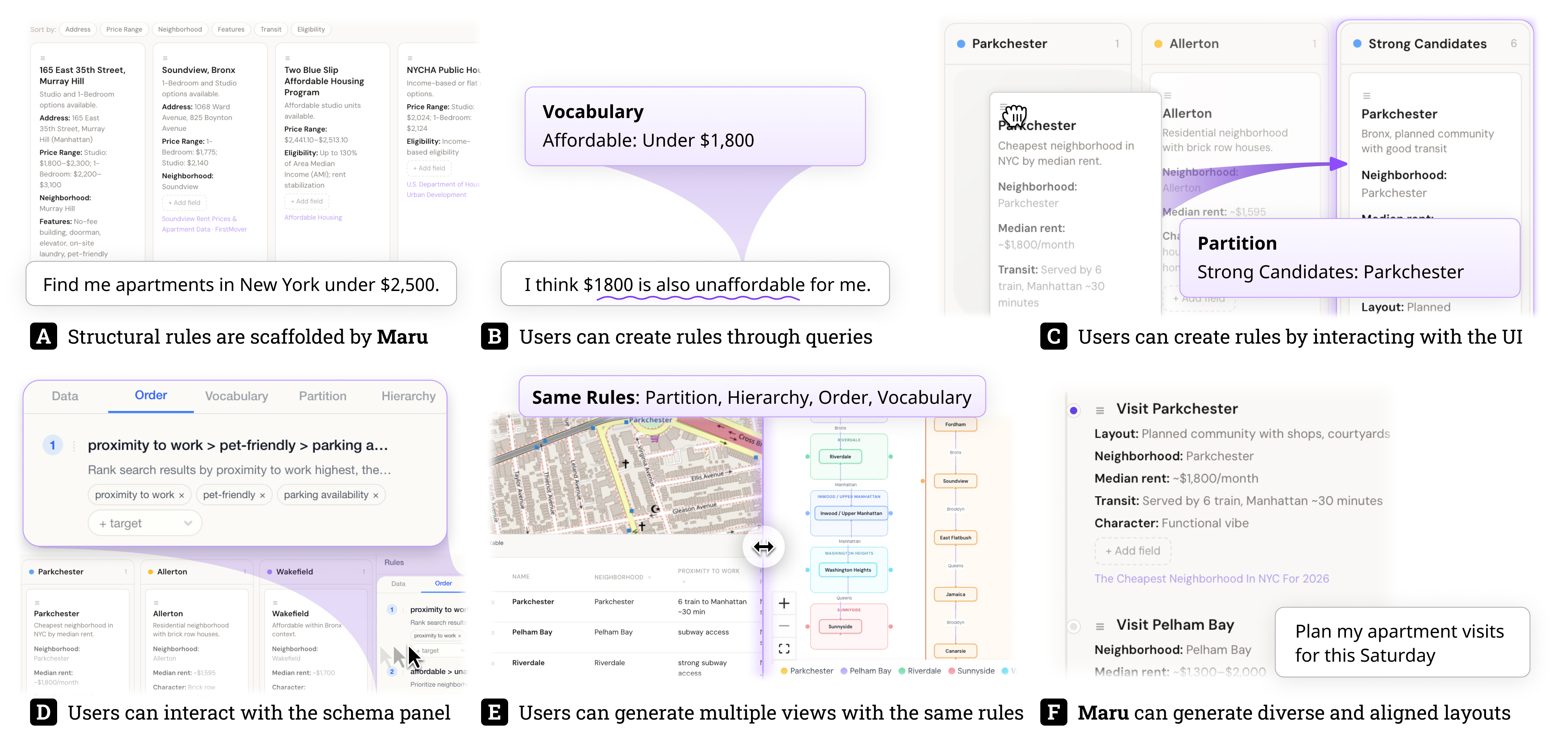}
    \caption{An envisioned scenario of a user searching for apartments in New York with \sysname{}. \textbf{(A)} Structural rules are scaffolded automatically from the initial query. \textbf{(B)} The user refines rules through natural language, adding a Vocabulary rule for "affordable." \textbf{(C)} \pointeight{The user drags an apartment into a new column within the generated UI itself, which updates a Partition rule without any explicit rule editing.} \textbf{(D)} \pointeight{Separately, the user adds an Order rule by editing the IA panel directly.} \textbf{(E)} The same accumulated rules produce multiple layout types without re-specification. \textbf{(F)} Rules accumulated across all three channels drive a task-specific Timeline layout for visit planning.}
    \Description{A six-panel scenario diagram showing a user searching for apartments in New York using Maru. (A) Structural rules are scaffolded by Maru: A card-grid layout displays four apartment listings with attributes such as address, price range, neighborhood, features, and transit. The query 'Find me apartments in New York under \$2,500' is shown below. (B) Users can create rules through queries: A tooltip callout reads 'Vocabulary — Affordable: Under \$1,800,' illustrating a vocabulary rule created from the query 'I think \$1800 is also unaffordable for me.' (C) Users can interact with the UI to rearrange rules: A kanban board shows three columns — Parkchester, Allerton, and Strong Candidates. A cursor drags a Parkchester card into the Strong Candidates column, implicitly updating the partition rule. The query reads 'Which neighborhoods should I focus on?' (D) Users can interact with the schema panel: The IA panel's Order tab is open, displaying an order rule 'proximity to work > pet-friendly > parking availability.' Below, a kanban view shows three neighborhood columns — Parkchester, Allerton, and Wakefield — with the Rules panel visible alongside. (E) Users can generate multiple views with the same rules: A banner reads 'Same Rules: Partition, Hierarchy, Order, Vocabulary.' Three layout types are shown side by side using identical rules — a map view, a ranked table with proximity-to-work data, and a node-link diagram of neighborhoods. (F) Maru can generate diverse and aligned layouts: A timeline-style list shows two entries — Visit Parkchester and Visit Pelham Bay — with structured details including layout description, neighborhood, median rent, transit, and character. A chat bubble reads 'Plan my apartment visits for this Saturday.
}
    \label{fig:scenario}
\end{figure*}

\username{} \revised{opens} \sysname{} to search for apartments in New York. She types \textit{"Find me apartments in New York under \$2,500"} and receives a card-grid. The system scaffolds two initial IA rules--a Partition grouping apartments by neighborhood and a Hierarchy nesting each apartment under it (Figure~\ref{fig:scenario}-\figlabel{A}). Noticing inconsistent use of "affordable", she types \textit{"Affordable means under \$1,800 and within 20 minutes of downtown,"} adding a Vocabulary rule applied to all subsequent generations (Figure~\ref{fig:scenario}-\figlabel{B}).

As her search deepens, \username{} asks \textit{"Which neighborhoods should I focus on?"} The system retrieves her Partition rules and selects a Kanban layout--each neighborhood becoming a column. She drags a Brooklyn apartment into a new column she labels \textit{"Strong Candidates."}  The drag updates her Partition rule, adding the new group without any explicit rule editing (Figure~\ref{fig:scenario}-\figlabel{C}). She then opens the IA panel directly and types an Order rule: \textit{"proximity to work > pet-friendly > parking availability,"} (Figure~\ref{fig:scenario}-\figlabel{D}). After a few iterations, her view has settled into a map alongside a table of listings. Wanting to understand how her neighborhoods and criteria relate to one another, she regenerates it as a diagram without changing any rules---the same partitions and hierarchy now render as a node-link graph (Figure~\ref{fig:scenario}-\figlabel{E}).

Finally, \username{} asks: \textit{"Plan my apartment visits for this Saturday."} The system generates a Timeline ordered by her priority rule, grouped by neighborhood partition, and annotated with her affordability vocabulary---a view that without accumulated IA context would produce a generic list with no awareness of her priorities, hierarchies, groupings, or definitions (Figure~\ref{fig:scenario}-\figlabel{F}).

\subsection{IA Construction}
\sysname{} is mainly composed of (1) an IA construction module and (2) an IA-based generation module. For IA construction, the user can manipulate the rules via chat, UI interaction or direct editing in the IA panel. When a session starts, \sysname{} detects the hierarchy and partition rules scaffolded from the generated UI as the starting point for the user's rules to build on. 

\subsubsection{Query-based Rule Detection}
The system detects IA rules from the user query. \revised{The user can submit a natural-language request to the system}, such as \textit{"I want to see the apartments by neighborhood, help me compare them"}. \revised{The rules are} inferred by an LLM, which uses the summarized chat context alongside the full active rule set, since rules can be applied on top of other rules as described in the framework.

\subsubsection{Interaction-based Detection}
Users can interact with the generated UI to communicate their structural preferences. Interactions are of two types: layout-agnostic and layout-specific, each mapping to specific IA elements according to the framework. An example of a layout-agnostic interaction is dragging a child component from one parent to another (e.g., a card from one tab to another), which triggers a partition rule update. A layout-specific interaction is connecting nodes with labeled links in a node-link diagram, which triggers a hierarchy rule. All supported layout types and their corresponding interactions are listed in Appendix~\ref{appendix:supported-interactions}. 

We enable interaction-based detection by rendering each layout component with built-in gesture handlers. Each component type exposes a defined set of interactions (e.g., drag-to-reorder, inline edit, lasso-group) that are captured as typed IA events and mapped to rule updates. 
\pointfour{All supported interactions update the rendered layout in place and write their IA rule directly, without invoking the generation pipeline. A new generation occurs only when the user issues a chat-level action---a new query, a reply, or an explicit regeneration request.}
\pointtwo{Extending \sysname{} to a new layout type involves defining the gestures that the component exposes and mapping each to one of the four IA elements. New layouts therefore require new gesture handlers but not new rule types, and the mapping in Table~\ref{tab:interactions} can grow without changing the framework.}

\subsubsection{The IA Panel}
The rules and data they are bound to are visualized and made accessible to the user through a panel. Each rule has a distinct visual representation, and the user can directly inspect, modify, and delete rules through the IA panel.

\subsection{IA-based Generation}

\begin{figure*}[h]
    \centering
    \includegraphics[width=0.9\linewidth]{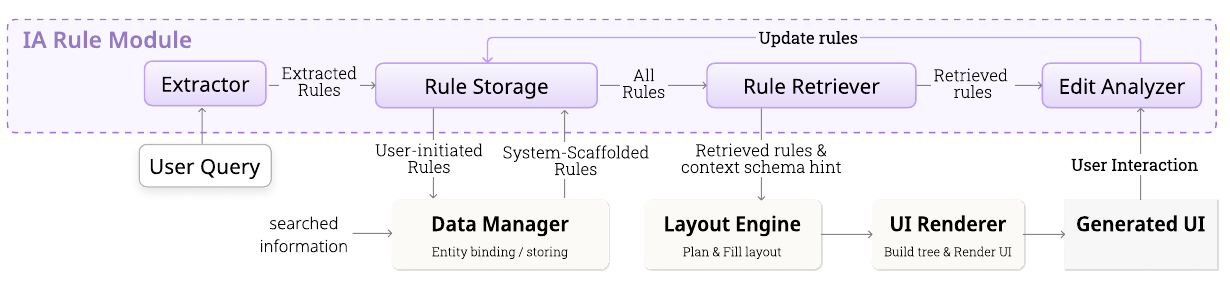}
    \caption{The \sysname{} pipeline. User queries are processed by the Extractor to detect IA rules, which are stored and retrieved alongside system-scaffolded rules to guide the Layout Engine. The UI Renderer builds and renders the generated interface, while the Edit Analyzer captures user interactions and feeds rule updates back into Rule Storage.}
    \Description{A pipeline diagram of the Maru system, divided into two layers. The top layer, enclosed in a purple dashed border and labeled 'IA Rule Module,' contains four components connected left to right: Extractor → Rule Storage → Rule Retriever → Edit Analyzer. The Extractor receives a User Query and outputs Extracted Rules to Rule Storage. Rule Storage holds both User-initiated Rules and System-Scaffolded Rules, and sends All Rules to the Rule Retriever. The Rule Retriever outputs Retrieved Rules to the Edit Analyzer, and also sends Retrieved rules and context schema hint downward to the Layout Engine. A curved arrow labeled 'Update rules' runs from the Edit Analyzer back to Rule Storage, completing the feedback loop. The bottom layer shows the generation pipeline: User Query sends searched information to the Data Manager (Entity binding / storing), which feeds into the Layout Engine (Plan & Fill layout), which feeds into the UI Renderer (Build tree & Render UI), which produces the Generated UI. User Interaction feeds upward into the Edit Analyzer.
}
    \label{fig:placeholder}
\end{figure*}

\subsubsection{Rule Retrieval}
Given a user prompt, the summarized chat context and the full active rule set are provided to the LLM to retrieve rules relevant to the current generation. For instance, if the user has partitions for "Neighborhood 1", "Neighborhood 2", and queries \textit{"Let's plan out the apartments in Neighborhood 1."}. The partition rule and its bound data entities for "Neighborhood 1" are retrieved. 

\subsubsection{Layout Selection}
UI generation proceeds in two steps: layout selection and layout filling. Layout selection takes as input the user query, few-shot examples of similar user needs and system views drawn from our literature codebook, and the retrieved hierarchy and partition rules. This step outputs a layout skeleton as a JSON structure specifying component types and group labels for the filling step. The user query is incorporated because some layouts carry task-specific priorities (e.g., a map view) that cannot be inferred from the rules alone. Since hierarchy and partition are the structural rules that determine grouping and nesting, these are used to guide layout selection. All supported layout types are listed in Appendix~\ref{appendix:supported-interactions} and an overview of the \sysname{} interface can be found in Appendix~\ref{appendix:system-layouts}. 

\subsubsection{Layout Filling}
Once the layout type is selected, the layout skeleton, the user query, and all four IA element rules with their bound data are sent to the LLM for content filling. Order and vocabulary rules participate in this stage because they are semantic rules that affect which items surface, how they are labeled, and how they are ranked within the structure. Both steps output JSON, which is parsed into typed, interactive block components for rendering.   

\subsection{Implementation Details}
\sysname{} is built with Electron, React, and TypeScript as a desktop client, with a main process handling all LLM calls and IA logic and a renderer process running the React UI. Rule detection, rule retrieval, layout planning, edit analysis, and response attribution use \texttt{gpt-4o-2024-11-20}. 

\paragraph{UI Rendering.}
Each assistant response is rendered through a two-stage pipeline. First, a layout planner selects a component type (e.g. card-grid, kanban, tabs, timeline) based on the query, retrieved data, and active IA rules. Second, a filler populates the chosen layout with content, producing a typed plan tree. The renderer walks this tree recursively, instantiating each node as an interactive React component with gesture handlers (drag-to-reorder, inline edit, delete, sort, lasso-group). Because the plan tree is a declarative data structure, the same response can be re-rendered or locally modified without re-invoking the LLM.

\paragraph{Entity Storing and Binding.}
Rules reference items via stable UUID pointers or loose concept labels. When data items are extracted from source content, concept references are automatically upgraded to UUID-based links, ensuring partition memberships and hierarchy relationships remain intact across item renames and field edits. 

%% file: sections/05_evaluation.tex
\section{User Study}

To evaluate whether IA-based persistence produces more aligned and personalized generative UIs, we conducted a comparative study with \sysname{} and a baseline system without the IA pipeline. We address three research questions:

\textbf{[RQ1] What are the benefits and costs of IA-based persistence for UI alignment?} \sysname{} accumulates IA rules across queries and interactions, applying them persistently on subsequent generations, while the baseline system does not. This distinction lets us isolate the effect of IA-based persistence: does carrying forward user-constructed structure improve alignment over the course of a session, or does accumulated rule state introduce its own costs?

\textbf{[RQ2] How do users communicate IA preferences, and what does their engagement reveal?} \sysname{} provides three channels through which users can communicate structural preferences. Understanding how rules are constructed in practice and how engagement varies across users reveals whether the shared IA language is accessible across different kinds of users.

\textbf{[RQ3] How does IA-based persistence bound the space of generated UIs?} IA rules translate user-constructed structure into concrete layout decisions. For IA rules to produce structurally distinct, user-personalized outputs, different rule states must yield meaningfully different UIs.


\subsection{Baseline System}
\pointthree{We designed the baseline as an ablation of the IA layer.} 
The baseline probe shared the same pipeline as \sysname{}, supporting the same UI interactions, layout types, and queries. However, the IA module was removed, with no rule detection, rule storage, codebook-derived schema hints, or rule-based layout examples. \pointthree{Both conditions received the same base context at each generation, including the current user query, the retrieved data items, a summary of the immediately preceding generation, and the user's full prior interaction history.} User interactions were instead logged semantically (e.g., \texttt{Moved item "Brooklyn Apt" to Tab "To Consider"}) and included in the prompt as prior interaction context. \pointthree{The two conditions therefore differed only in how that history was represented.} In \sysname{}, the same interaction would be encoded as a typed Partition rule; in the baseline, it remains an unstructured log entry. The prompt differences are explained in detail in Appendix~\ref{tab:baseline-comparison}.

\subsection{Participants, Procedure, and Measurements}
We recruited participants through an online community at our institution. A total of 12 participants (average age = 24.3, 5 male, 7 female) were recruited. All participants reported daily or near-daily LLM use and prior experience with information-intensive tasks. 

Each participant completed three task sessions (S1-3). In the first two sessions (S1 and S2), participants completed two assigned tasks: graduate school program or career exploration and picnic planning. Each task was completed with one system--\sysname{} or the baseline--with system order counterbalanced across participants. In the third session (S3), participants used \sysname{} on a real information task they had prepared in advance. Full session procedures and detailed task descriptions are provided in Appendix~\ref{tab:s3-tasks}. Each participant was compensated approximately \$20 USD, and the study was conducted under institutional IRB approval.

The system logged timestamped records of all participant actions (queries, UI interactions, and rule edits) and system outputs (LLM prompts, responses, chosen layouts, and generation durations) to per-session log files. Session metadata including participant ID, condition, and duration was tracked in a separate log file for cross-session analysis.

%% file: sections/06_results.tex
\section{User Study Results}
Across 36 sessions, participants generated 220 UIs and performed 585 interactions, constructing a total of 1,292 IA rules (838 user-attributable and 454 system scaffolded). Sessions spanned \revised{nine of the supported layout types}, with participants in the \sysname{} condition producing layouts across an average of 2.7 distinct types per session. \pointthree{Prompt lengths were comparable across conditions (median 3.4k tokens per model call in the baseline vs.\ 3.3k in \sysname{}), indicating the conditions differed in how prior context was represented rather than how much was provided.} We report findings across three RQs: how IA-based persistence shapes UI alignment, how users construct IA rules across interaction channels, and how accumulated rule states produce personalized UI trajectories. We provide a gallery of all generated widgets as supplementary material.


\subsection{RQ1. What are the benefits and costs of IA-based persistence for UI alignment?}



\paragraph{\sysname{}'s IA-based persistence produces more aligned UIs across the session.} 

After each generation, participants rated UI alignment with their needs on a binary scale (accept/reject). \pointfive{\sysname{}'s UIs were more likely to be accepted overall~(OR $= 2.08$, 95\% CI $[1.25, 3.47]$, $p = .005$),} \pointthree{and this advantage was not attributable to context volume---the effect persisted after statistically adjusting for the prompt tokens consumed by each generation (adjusted OR $= 2.40$, 95\% CI $[1.28, 4.47]$, $p = .006$).} In the baseline condition, this approval rate started at 71\% in the first half of the session but collapsed to 33\% in the second half\pointfive{~(OR $= 0.19$, 95\% CI $[0.08, 0.45]$, $p < .001$)}.
In contrast, \sysname{}-generated UIs started at 74\% and held at 61\% throughout the second half\pointfive{~(OR $= 0.63$, 95\% CI $[0.15, 2.65]$, $p = .53$)}. This stability is reflected in generation efficiency as well: \pointfive{participants reached a satisfactory output in fewer generations under \sysname{}~($M = 5.58$, $SD = 1.62$) than the baseline~($M = 7.75$, $SD = 2.09$; Wilcoxon signed-rank $V = 0$, $p = .008$).} When asked which condition felt more aligned with their preferences, 9 participants preferred \sysname{}. Participants attributed their preference to the system's ability to carry forward what they had established: P8 noted that they could "see effort in the results---it was persistent with what I had valued before...the system understood what I was trying to convey." 


\paragraph{Participants leveraged IA rules to align the UI to their needs.}
Without IA-based persistence, baseline users developed manual compensatory strategies to maintain structural continuity. This is visible in query behavior: in \sysname{}, the average query length of user prompts declined by 10.6 words (28.7 → 18.1) from the first half to the second half of the session, as accumulated rules absorbed the specificity that users would \revised{otherwise} have to re-articulate. In the baseline, the decline was only 2.5 words (21.8 → 19.3), with some participants copy-pasting previous queries and appending new conditions to compensate for the lack of carry-over. Others avoided generating new outputs altogether, returning to a previous UI and continuing to interact within it, since the state of a single widget persisted internally even if it would not carry over to the next generation. P2 illustrates the failure that drove this: "When I ask the baseline system to add the activities into the calendar that it had created before, it just creates a new calendar and ignores the calendar that it created." 
Both strategies---re-articulating their preferences or limiting their interactions to a single output---reflect the same underlying problem: without a persistent structural layer, users had no basis to expect that a new generation would preserve what they had already established. In \sysname{}, this overhead largely disappeared, and short refinements like "show me more options" replace multi-sentence specifications as rules carried structural context forward.

\paragraph{IA rules hindered alignment when persisting across sub-tasks during context shifts.}
While IA-based persistence improved alignment overall, it degraded in two conditions. The first was sub-task transitions. When participants shifted focus within a session, rules from the prior sub-task were retrieved and applied to the new context. Generations where cross-topic rules were retrieved had a 23\% thumbs-down rate compared to 18\% for on-topic retrievals. P5 shifted from PhD program search to HCI lab exploration and rated three consecutive UIs as unsatisfactory, each carrying 5–6 rules from the cost-of-living sub-task---"Housing" partitions and "total" ordering that had no relevance to lab selection. P11 similarly observed an "Off-campus" partition from venue search leak into food recommendations during picnic planning. In each case, the system correctly persisted rules that had been valid in their original context but failed to recognize that the user had moved on. Users found this frustrating, with P9 saying that "I didn't like that something from a separate task carried over into the next." This suggests that IA persistence in generative UI systems needs to take in consideration a task-aware scoping~\cite{cao2025generative} that tracks which rules belong to which sub-task context to help scope retrieval appropriately.

\paragraph{IA rules hindered alignment when over-accumulated.} 

Rule accumulation was another condition where IA-based persistence degraded alignment. Alignment suffered when active rule counts rose continuously across sessions: from a mean of 15.2 in the first quarter to 37.5, 59.1, and 101.6 in subsequent quarters. This growth was driven by two mechanisms. First, the system's scaffolding mechanism generated new structural rules with each generation by scanning the data schema of the generated plan---an average of approximately 24 system-scaffolded hierarchy rules per generation that were added to the rule state without user action or awareness. Second, user interactions accumulated refinement rules that were never pruned, merged, or expired, even when they were no longer relevant. Although retrieval matched queries against the active rule set, \sysname{} lacked a mechanism to identify rules the user had mentally discarded without explicit pruning. The two participants with the highest rule counts---P5 (194 rules, 33\% approval) and P11 (198 rules, 50\% approval)---had the lowest approval ratings in the study, and the slight second-half decline (74\% → 61\%) in \sysname{}'s approval rate is partially attributable to this accumulation. This suggests that both user-created and system-scaffolded rules need context-sensitive lifespans, with relevance weighting that accounts for recency.

\subsection{RQ2. How do users communicate IA preferences, and what does their engagement reveal?}




\begin{figure*}[h]
    \centering
    \includegraphics[width=\linewidth]{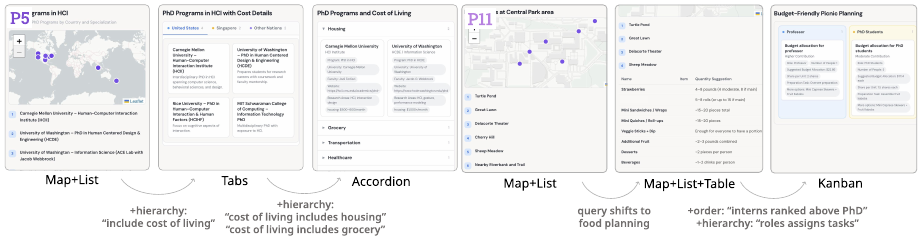}
    \caption{UI layout trajectories for P5 (grad school search) and P11 (picnic planning) illustrating how IA rules shape layout changes differently across users. P5's accumulating hierarchy rules trigger progressive layout shifts (Map+List → Tabs → Accordion), each transition reflecting the growing parent–child structure in their rule state. P11's rules accumulate across queries without producing visible layout changes until a new hierarchy and order rule combination triggers a shift to kanban.}
    \Description{A side-by-side layout trajectory diagram for two participants, P5 and P11, each showing three sequential UI layouts driven by accumulating IA rules. P5 (grad school search): The first layout is a Map+List view showing HCI PhD programs plotted on a world map with a numbered list below (e.g., Carnegie Mellon, University of Washington). A hierarchy rule '+hierarchy: include cost of living' triggers a transition to a Tabs layout, showing PhD programs grouped by country (United States, Singapore, Other Nations) with program details in cards. A second hierarchy rule '+hierarchy: cost of living includes housing, cost of living includes grocery' triggers a transition to an Accordion layout, where cost-of-living categories (Housing, Grocery, Transportation, Healthcare) appear as collapsible sections under each university. P11 (picnic planning): The first layout is a Map+List view showing picnic locations near Central Park (Turtle Pond, Great Lawn, Delacorte Theater, etc.) alongside a numbered list. The query shifts to food planning, producing a Map+List+Table layout that adds a food quantity table (items like Strawberries, Mini Sandwiches, Veggie Sticks) alongside the map. New rules '+order: interns ranked above PhD' and '+hierarchy: roles assigns tasks' trigger a transition to a Kanban layout, with two columns — Professor and PhD Students — each containing budget allocation and preparation task details.
}
    \label{fig:trajectories}
\end{figure*}

\begin{table}[t]
\centering
\small
\begin{tabular}{lrrrr}
\toprule
 & Partition & Hierarchy & Order & Vocabulary \\
\midrule
UI interaction & 305 & 149 & 53 & 132 \\
Query extraction & 54 & 65 & 19 & 32 \\
Manual panel edit & 5 & 5 & 9 & 10 \\
\midrule
Total (838) & 364 & 219 & 81 & 174 \\
\bottomrule
\end{tabular}
\caption{Distribution of the 838 user-attributable IA rules by extraction method and IA element.}
\Description{A table showing the distribution of 838 user-attributable IA rules by extraction method and IA dimension. Columns represent the four IA dimensions: Partition, Hierarchy, Order, and Vocabulary. Rows represent three rule creation methods: UI interaction (305, 149, 53, 132), Query extraction (54, 65, 19, 32), and Manual panel edit (5, 5, 9, 10). Row totals are 364 Partition, 219 Hierarchy, 81 Order, and 174 Vocabulary, summing to 838. UI interaction is the dominant source across all four dimensions.
}
\label{tab:ia_rule_distribution}
\end{table}


\paragraph{IA rules emerge from a collaborative loop of system scaffolding and user-expressed preferences.}
Across all sessions, \sysname{}'s combination of scaffolding and user refinement produced 1,292 active rules, of which 838 were user-attributable, created or modified through interactions, queries, or manual panel edits, and 454 were system-inferred rules used for layout population. When the system surfaced rules inferred from user behavior, participants largely agreed: of 94 rules that received explicit feedback, 88 were accepted \pointfive{(94\%, 95\% CI $[87\%, 97\%]$)}. This suggests that the IA framework functions as a genuinely shared language, and the structural inferences the system makes from user behavior are ones users recognize as their own. 

\paragraph{Users rarely created new IA rules, but frequently interacted with the ones they were given. }
Of the 838 user-attributable rules, 76\% originated from UI interactions, 20\% from natural language extraction of queries, and only 3\% from manual panel editing (Table~\ref{tab:ia_rule_distribution})---meaning 97\% of user-created rules were built without any deliberate engagement with the IA schema. Only 29 rules across all 12 participants were manually created or modified through the panel. The interaction-to-rule mapping was the main channel through which users communicated their preferences: sorting items generated order rules, removing or adding fields generated hierarchy rules, deleting items generated partition rules, and lasso-grouping generated both partition and vocabulary rules. The reliability of this mapping suggests that interaction behavior is the most natural channel through which users express structural preferences---and therefore future IA-based generative UI systems should be designed to capture a broad range of interaction types as structural signals.

\paragraph{Users produced and modified unintuitive IA rules through direct manipulation, while articulating intuitive ones through prompts}
Although users found semantic rules (vocabulary, order) to be the most intuitive dimensions, they overwhelmingly generated, interacted with, and modified structural ones (partition, hierarchy). 
In post-session interviews, participants ranked vocabulary as the most natural and immediately understandable, followed by order, partition, and hierarchy---which multiple participants described as difficult to grasp (P5-7, P11). However, our behavioral data revealed that partition was the most common user-created rule type at 43\% (364/838), followed by hierarchy at 26\% (219), vocabulary at 21\% (174), and order at 10\% (81). UI interactions produced a balanced spread across all four types (21\% vocabulary, 8\% order, 48\% partition, 23\% hierarchy), while natural language extraction skewed toward hierarchy (38\%) as users mentioned fields in queries. \pointfive{Participants' accounts suggest this ordering followed how much each element matched concepts they already had words for. Sorting and defining a term are operations users bring from everyday software, while hierarchy had no such counterpart. Participants performed hierarchy operations constantly, but adding or removing a field read to them as adjusting what they wanted to see rather than editing a structural relation. Partition fell in between, with P4 and P6 both reporting that they only came to understand how a partition worked after seeing one applied.}
This finding extends prior knowledge about the value of both prompting and direct manipulation in AI systems \cite{masson2024directgpt}. 
\pointfive{Users engaged with IA structurally through their behavior even when they could not articulate it, suggesting that interaction is the primary channel through which the shared language operates.}

\paragraph{Users without design expertise increasingly customized the UI with IA rules throughout their task.}


Although \sysname{} added persistence to what users expressed, it did not change how control felt: perceived control was nearly identical across conditions (\sysname{}: 3.25/5, Baseline: 3.00/5, $\Delta$=+0.25, p=.555), since the main control channel itself (interaction and query) was the same in both conditions. What differed was engagement depth. Participants without design experience rarely opened the IA panel, interacting through queries and direct manipulation alone. Yet over the course of their sessions, these participants became more intentional in how they communicated with the system. P3 noted the system was "aligning well when I ask narrowing questions," P4 described shifting from exploration to deliberate comparison, and P10 began leveraging persistence intentionally: "Because I asked to include safety, it now always includes it." Design-experienced participants engaged more explicitly from the outset: P5 prompted with IA terminology directly, P12 requested specific layout types, and P11 remarked they would rather build the interface themselves. These patterns suggest users engaged with the IA layer at different depths depending on their design expertise, and that the IA framework---even when operating invisibly---can serve as an entry point through which users with no design background gradually develop structural agency over their generated UIs.

\subsection{RQ3. How does IA-based persistence bound the space of generated UIs?}

\paragraph{IA rules bound layout divergence across users.}
Twelve participants each performed two assigned tasks---career/grad school finding and picnic planning---one in each condition. In both tasks and both conditions, first-generation layouts showed approximately 50\% agreement, as both systems made comparable initial guesses based on the query alone. Over the course of each session, however, the conditions diverged. 
In the career task, all six baseline participants' final layouts converged to tables. Without persistent structural preferences to differentiate on, every participant ended at that default regardless of their individual needs. In \sysname{}, the six career participants ended on four distinct layout types---maps, accordion, tabs, and list. The picnic task followed the same directional pattern: \sysname{} produced six distinct final layout types compared to the four in the baseline. The career task converged more strongly because it has a clearer default representation: picnic planning, with its multiple facets (venue, food, schedule, budget), lacks a single layout, so both conditions produced more variety. 

\paragraph{IA elements route to distinct layout types, and rule changes \pointseven{precede} layout transitions.}
Cross-tabulating active rule states with generated layouts across all system sessions reveals systematic mappings between IA dimensions and layout types. When partition rules were present, 33\% of generated layouts used grouped representations---kanban views, tabbed interfaces, or accordions---compared to 0\% without. When five or more hierarchy rules had accumulated, 10\% of layouts were tables, compared to 0\% with fewer. Order rules were associated with a 17\% rate of table or list layouts versus 5\% without. Notably, these \revised{mappings} reflect layout type rather than complexity, since average section count remained between 1.2 and 1.6 regardless of how many rules were active. 
\pointseven{These rule changes preceded and plausibly drove the layout transitions that followed: of 45 layout transitions observed across \sysname{} sessions, 29 (64\%) showed a clear link between a newly created rule and the subsequent layout change. Partition additions were followed by a shift to grouped layouts in 16 cases; order rule additions preceded transitions to sequenced or ranked layouts in 7 cases, such as P1's shift to a ranked list after creating an order rule and P9's shift to a timeline after expressing temporal ordering preferences; hierarchy accumulation preceded table layouts in 6 cases as growing field-level rules made tabular representations increasingly appropriate.}

\paragraph{Individual rule profiles produce distinct UI trajectories within the same task in \sysname{}.}
The personalization effect is most visible when comparing participants who performed the same assigned task. Among the six picnic planners in the \sysname{} condition, each user's layout trajectory was shaped by their individual rule profile: P4's temporal order rules (\textit{"order lunch first"}) produced timeline layouts, while P9's partition rules across snacks, drinks, and activities produced accordion and tabbed layouts. The divergence is sharpest among the three participants who independently planned Barcelona trips in S3. P6, with rich rules across all four IA elements (8 vocabulary, 5 order, 20 partition, 55 hierarchy), progressed through accordion, diagram, and timeline layouts. P11, with minimal interaction and no vocabulary or order rules (38 total), remained on map+tabs layout for five consecutive generations. P12, whose interactions were predominantly deletions creating partition exclusion rules (24 total), converged to simplified tabs. 
Across all system sessions, this relationship between rule richness and layout diversity was consistent: participants with fewer than 20 rules experienced an average of 1.5 distinct layout types, rising to 2.3 types with 20-50 rules and 3.0 with 50 or more. \pointseven{This suggests that different rule states produce meaningfully different output spaces, and that accumulated structural intent can bound which layouts a GenUI system arrives at rather than converging on a single optimal one.}

%% file: sections/07_discussion.tex

\section{Discussion}

\subsection{Designing for Persistence Across Levels and Contexts in GenUI}
Our results suggest that persistence is a key factor in making user-aligned GenUI for information tasks; however, we address only IA-level structural persistence. Yet UIs have many more semantic and visual layers, and our findings hint that different users want different layers persisted. Non-expert users were largely satisfied with IA-level persistence, while users with more design experience  sought control at finer granularities like layout type and component structure. This points toward a layered model of persistence for future GenUI systems: from high-level structural rules accessible to all users, down to component-level handles for users with the vocabulary to use them.

\pointsix{If levels concern how much of the UI persists, scope concerns how long and where.} Our findings show that unbounded rules accumulated beyond their useful context by bleeding across sub-task boundaries and persisting long after users had moved on. \pointsix{What our findings point to is not less persistence but a boundary on it.} Future systems need mechanisms that let rules expire or be scoped to sub-task contexts. \pointsix{A task-schema~\cite{cao2025generative} could determine whether a rule belongs to the current sub-task and should be carried into it at all, while a decay mechanism over accumulated context~\cite{shaikh2025creating} could down-weight rules the user has stopped acting on.} \pointthree{Alongside these, systems that surface what the model has assumed~\cite{zamfirescu2025pail, gebreegziabher2025mocha} could elicit preferences the user never thought to state.}

\pointseven{Beyond scoping within a session, a natural extension is reuse across instances of a task. In exploring this, we found the rules as we designed them were difficult to carry across tasks without more observation of the user and an explicit mechanism for generalization. Their concreteness is what lets them determine specific layout and content decisions, but it also ties them to the task they were formed in. What survived abstraction was a general sense of what the user cared about rather than something that could map to a UI decision. One direction may be to treat abstracted user models as a basis for IA rules, where approaches that synthesize recurring behavior across sessions~\cite{zhao2026behavior} supply the higher-level structure from which task-specific rules are formed.}

\subsection{IA as a Shared Language for End-User Customization}
It has long been observed that people rarely customize their software due to high interaction costs and the requirement for technical \revised{expertise}~\cite{nardi1993small}. GenUI begins to shift this dynamic by replacing programming with natural language and interaction as the medium for customization~\cite{kim2022stylette,min2026gradual}. Our findings suggest that IA rules represent a promising layer for this shift---with 97\% of rules built without deliberate schema engagement, the IA layer absorbs customization intent from natural interaction rather than requiring users to engage in architectural modeling.
\pointfive{Hierarchy illustrates this most clearly. It was the element participants found hardest to grasp, yet it accounted for the second-largest share of user-created rules---users produced structure they could not name. Rather than teaching users a schema, then, systems can let them act through operations they already recognize and infer the structure from what they do.}

Throughout the study, we observed a behavioral shift. Users gradually learned to communicate more deliberately with \sysname{} by specifying preferences explicitly once they saw the system would respond. Over time, users shifted to more targeted refinements, trusting the system to carry forward established preferences as accumulated rules handled more of the structure. 
\pointseven{Within our sessions, this suggests IA rules can operate as a shared language users grow more comfortable working through, though how such a shift develops over longer use remains open.}


\subsection{Establishing Meaningful Boundaries in Non-Deterministic GenUI}

Our findings suggest GenUI research should shift from identifying the single optimal layout for a query toward ensuring the generative design space is shaped by the user's preferences.
The trip comparison shows this well, as users doing the same task produced their own distinct and valid UI trajectories through \sysname{}. This points to a broader shift in how we think about GenUI. Rather than asking "What is the best layout for this task?" the more useful question is "What conditions make a layout satisfying for this user?" IA rules represent one way to establish those conditions by guiding the system toward outputs that fit within the user's accumulated structural intent. The non-deterministic nature of generation therefore becomes less something to eliminate and more something to structure, since variety can be acceptable and even desirable as long as outputs stay within meaningful boundaries. \pointsix{This also suggests that generation becomes more valuable later in a task than at the start. The difference between \sysname{} and the baseline was largest in the second half of sessions, once users had built up structure of their own. Early on, a well-designed default may work just as well, and IA-guided personalization matters more as a user's needs move away from it.} What remains open is how to design systems that help users establish and refine those boundaries over time, and how to ensure that the range of outputs the system can produce is visible enough that users can steer it intentionally.

\subsection{Towards the GenUI ``Stack''}
Software applications are built on stacks---layered technologies each handling a distinct concern. We envision a similar stack for GenUI will also begin to take shape: task-driven data models structure what information to surface for a given activity \cite{cao2025generative, lam2025just}, UI specification languages provide a shared grammar for describing interface intent \cite{google2025a2ui, beaudouin2026belidor}, and malleable interface frameworks determine how generated outputs can be customized \cite{min2025malleable, min2025meridian}. IA sits between these layers, translating the structural logic users build during a task into concrete generation decisions and persisting it across the session. \pointtwo{The stack must also span activities, not only layers, as structural abstractions of this kind appear in generative systems for other domains~\cite{aveni2025generative, angert2023spellburst}. Participants who took \sysname{} beyond sensemaking suggest the four elements can hold as a foundation while their relative importance shifts across activities.} The broader challenge of building GenUI systems aligned with individual users over time will require the community to develop and connect these layers together. We hope IA persistence serves as a concrete starting point for enriching the tooling around GenUI through multiple layers. 

%% file: sections/99_appendix.tex
\captionsetup{width=.8\textwidth}
\clearpage
\onecolumn

\section*{Appendices}
\renewcommand{\thesection}{A.\arabic{section}}
\renewcommand{\thesubsection}{\thesection.\arabic{subsection}}
\renewcommand{\thefigure}{A.\arabic{figure}}
\renewcommand{\thetable}{A.\arabic{table}}
\setcounter{section}{0}
\setcounter{figure}{0}
\setcounter{table}{0}
\setcounter{page}{1}

\renewcommand{\arraystretch}{1.4}
\section{Literature Coding Process}

\subsection{Full Set of Literature Papers for Each Round}
\begin{table}[H]
\centering
\caption{Literature Coding Process Across Rounds}
\Description{A table showing the literature coding process across three rounds, with columns for Round, Sensemaking Theory, and Interactive Systems. In the 1st round, Sensemaking Theory papers are [52, 53, 74, 80] and Interactive Systems papers are [10–13, 27, 49, 50, 87]. In the 2nd round, Sensemaking Theory is [38] and Interactive Systems are [41, 47, 48, 54, 67, 73, 98]. In the 3rd round, Sensemaking Theory papers are [31, 104] and Interactive Systems are [2, 14, 17–20, 28, 29, 33, 68, 75, 76, 79, 86, 93–95, 97, 101, 103].
}
\label{tab:literature-set}
\small
\begin{tabularx}{\linewidth}{p{0.08\linewidth}XX}
\toprule
\textbf{Round} & \textbf{Sensemaking Theory} & \textbf{Interactive Systems} \\
\midrule
1st & \cite{russell1993cost, ma2023browsing, marchionini2006exploratory, rieh2016towards} & \cite{chang2021tab, chang2019searchlens, chang2020mesh, hahn2018bento, liu2024selenite, liu2022wigglite, chang2021tabs, suh2023sensecape} \\
2nd & \cite{kittur2013costs} & \cite{yen2024memolet, marchionini2003towards, liu2019unakite, rachatasumrit2021forsense, liu2022crystalline, palani2022interweave, kuznetsov2022fuse} \\
3rd & \cite{hoeber2025design, zhu2024patterns} & \cite{ward2021orgbox, hearst2013sewing, zhang2008citesense, ahn2010you, subramonyam2020texsketch, dontcheva2006summarizing, vermette2015cheatsheet, ruotsalo2014intentradar, fogarty2008cueflik, robertson1998data, wang2025intentprism, xu2025duetui, han2020designing, fok2025facets, kang2023synergi, chang2023citesee, park2023foundwright, zheng2024disciplink, ros2024textdata, fok2024marco} \\
\bottomrule
\end{tabularx}
\end{table}


\subsection{Finalized Literature Codebook}
\begin{table}[H]
\centering
\caption{Full mapping of IA elements to user needs, UI components, and interactions across coded systems.}
\Description{A table mapping IA elements to user needs, UI components, and interactions across coded systems. It has five columns: Category, Element, User Needs, UI Components, and Interactions. Under the Structural category, the Partition element addresses three user needs: grouping related items laterally for visual comparison, regrouping flexibly as mental model evolves, and keeping interest facets distinct and independently adjustable. UI components include spatial clusters, tab groups and card clusters, color coding, and faceted sections. Interactions include drag-and-drop into groups, tag and label assignment, and spatial arrangement. The Hierarchy element addresses three user needs: decomposing complex tasks into nested subtasks, structuring evidence under options, and maintaining provenance of collected information. UI components include tables, accordions and toggles, nested cards, and node-link diagrams. Interactions include drag-and-drop into containers, context menus for nesting, and inline role assignment. Under the Semantic category, the Order element addresses three user needs: surfacing high-priority items to direct attention, weighting criteria by relative importance to reflect current priorities, and continuously adjusting relative importance across rules. UI components include spatial placement, weight badges and size emphasis, threshold filtering and collapsing, and pin and star actions. Interactions include drag-and-drop reordering, weight sliders, and priority selectors. The Vocabulary element addresses three user needs: anchoring comparison structures in personal terminology, evaluating options consistently using self-defined criteria, and persisting personal terms as navigational labels across sessions. UI components include persistent labels and column headers, tag chips and filter labels, and tooltips and inline highlights. Interactions include inline text input, tag creation, and term extraction from content. All entries include citation references.
}
\label{tab:literature-codebook}
\small
\begin{tabularx}{\linewidth}{p{0.10\linewidth}p{0.10\linewidth}XXX}
\toprule
\textbf{Category} & \textbf{Element} & \textbf{User Needs} & \textbf{UI Components} & \textbf{Interactions} \\
\midrule
\multirow{4}{*}{Structural} & \multirow{4}{*}{Partition}
& Group related items laterally for visual comparison & Spatial clusters~\cite{suh2023sensecape,yen2024memolet,rachatasumrit2021forsense,palani2022interweave,ward2021orgbox,hearst2013sewing,ahn2010you,robertson1998data,park2023foundwright,hoeber2025design} & Drag-and-drop into groups~\cite{rachatasumrit2021forsense,kuznetsov2022fuse,chang2021tabs} \\
& & Regroup flexibly as mental model evolves & Tab groups / card clusters~\cite{chang2021tab,chang2021tabs,xu2025duetui,ros2024textdata,kuznetsov2022fuse,zhang2008citesense,hoeber2025design,ma2023browsing,fok2024marco} & Tag / label assignment~\cite{chang2019searchlens,fogarty2008cueflik} \\
& & Keep interest facets distinct and independently adjustable & Color coding~\cite{chang2021tab,yen2024memolet,rachatasumrit2021forsense,chang2023citesee,hoeber2025design} & Spatial arrangement~\cite{rachatasumrit2021forsense,suh2023sensecape} \\
& & & Faceted sections~\cite{chang2021tab,marchionini2006exploratory,chang2019searchlens,chang2021tabs,fogarty2008cueflik,hoeber2025design,marchionini2003towards} & \\
\midrule
& \multirow{4}{*}{Hierarchy}
& Decompose complex tasks into nested subtasks & Tables~\cite{chang2020mesh,liu2019unakite,liu2022crystalline,fok2025facets} & Drag-and-drop into containers~\cite{liu2019unakite,kuznetsov2022fuse,chang2021tabs} \\
& & Structure evidence under options & Accordions / toggles~\cite{chang2021tabs,kuznetsov2022fuse,kang2023synergi} & Context menus for nesting~\cite{chang2021tabs,suh2023sensecape} \\
& & Maintain provenance of collected information & Nested cards~\cite{kuznetsov2022fuse,xu2025duetui,hoeber2025design,ward2021orgbox} & Inline role assignment~\cite{liu2019unakite,liu2022crystalline} \\
& & & Node-link diagrams~\cite{yen2024memolet,palani2022interweave,subramonyam2020texsketch,wang2025intentprism,russell1993cost,han2020designing} & \\
\midrule
\multirow{4}{*}{Semantic} & \multirow{4}{*}{Order}
& Surface high-priority items to direct attention & Spatial placement~\cite{hahn2018bento,liu2024selenite,chang2021tabs,liu2019unakite,liu2022crystalline,kuznetsov2022fuse,hearst2013sewing,park2023foundwright} & Drag-and-drop reordering~\cite{chang2020mesh,liu2019unakite,kuznetsov2022fuse,ruotsalo2014intentradar} \\
& & Weight criteria by relative importance to reflect current priorities & Weight badges / size emphasis~\cite{chang2019searchlens,chang2020mesh,ruotsalo2014intentradar,fogarty2008cueflik,robertson1998data,chang2023citesee} & Weight sliders~\cite{fogarty2008cueflik,chang2020mesh} \\
& & Continuously adjust relative importance across rules & Threshold filtering / collapsing~\cite{liu2022wigglite,chang2020mesh,chang2021tabs} & Priority selectors~\cite{chang2021tabs,hahn2018bento} \\
& & & & Pin / star actions~\cite{hahn2018bento,liu2022wigglite,liu2022crystalline,vermette2015cheatsheet} \\
\midrule
& \multirow{3}{*}{Vocabulary}
& Anchor comparison structures in personal terminology & Persistent labels / column headers~\cite{chang2020mesh,kittur2013costs,liu2019unakite,liu2022crystalline,hearst2013sewing,fok2025facets,park2023foundwright,marchionini2003towards} & Inline text input~\cite{chang2020mesh,liu2024selenite,kittur2013costs,liu2019unakite,liu2022crystalline} \\
& & Evaluate options consistently using self-defined criteria & Tag chips / filter labels~\cite{liu2024selenite,vermette2015cheatsheet,zheng2024disciplink,liu2022wigglite} & Tag creation~\cite{liu2022wigglite,ros2024textdata,hearst2013sewing,chang2023citesee,rieh2016towards,zhu2024patterns} \\
& & Persist personal terms as navigational labels across sessions & Tooltips / inline highlights~\cite{dontcheva2006summarizing,liu2024selenite,chang2023citesee,ros2024textdata,zhang2008citesense,zheng2024disciplink,vermette2015cheatsheet} & Term extraction from content~\cite{subramonyam2020texsketch,suh2023sensecape,palani2022interweave,dontcheva2006summarizing,fogarty2008cueflik,han2020designing} \\
\bottomrule
\end{tabularx}
\end{table}

\clearpage
\section{\sysname{} Details}

\subsection{\sysname{} Interface}
\label{appendix:system-layouts}

\begin{figure}[h]
    \centering
    \includegraphics[width=\textwidth]{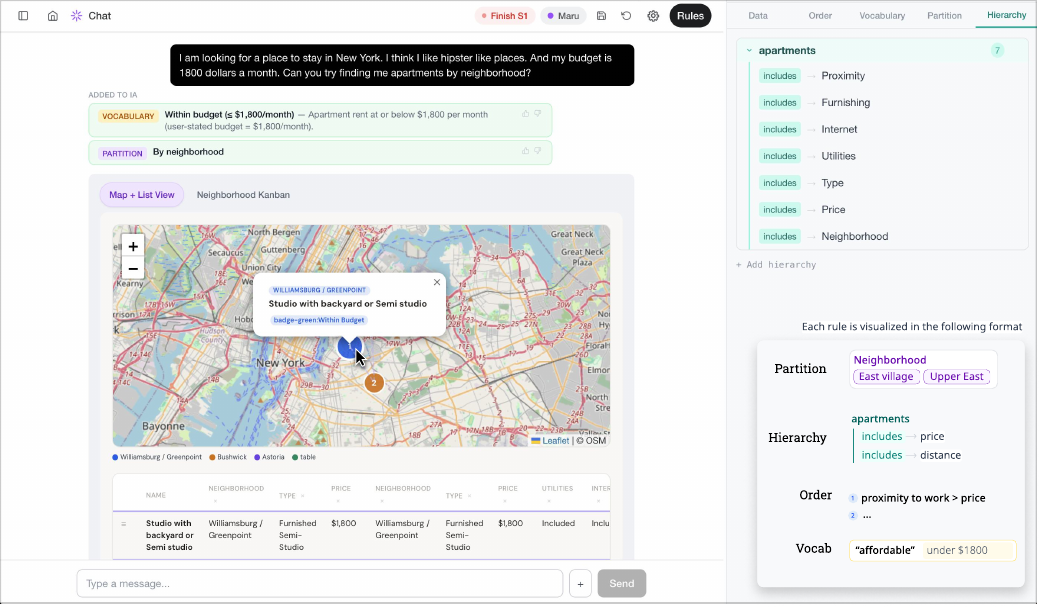}
    \captionsetup{width=\textwidth}
    \caption{Overview of the \sysname{} interface. \textbf{(1) Chat Panel (left):} Users 
    submit natural language queries and receive generated UI outputs. Rules 
    detected from queries and interactions are surfaced inline as tagged 
    labels (e.g., \textsc{Vocabulary}, \textsc{Partition}). \textbf{(2) 
    Generated UI (center):} The system renders interactive layout components 
    --- here, a Map + List View --- populated according to the user's active 
    IA rules. \textbf{(3) IA Panel (right):} The accumulated rule state is 
    visualized across five tabs (Data, Order, Vocabulary, Partition, 
    Hierarchy). Each rule is displayed in a consistent format showing rule 
    type, bound entities, and values, and can be directly inspected, 
    modified, or deleted.}
    \Description{A screenshot of the Maru interface with three panels. Left — Chat Panel: A top navigation bar shows buttons labeled 'Finish S1,' 'Maru,' and 'Rules.' A user message reads 'I am looking for a place to stay in New York. I think I like hipster like places. And my budget is 1800 dollars a month. Can you try finding me apartments by neighborhood?' Below, two detected rules are surfaced inline: a Vocabulary rule labeled 'Within budget (≤ \$1,800/month) — Apartment rent at or below \$1,800 per month' and a Partition rule labeled 'By neighborhood.' A text input field at the bottom reads 'Type a message...' Center — Generated UI: Two layout tabs are shown: 'Map + List View' (active) and 'Neighborhood Kanban.' The map displays New York City with numbered pins for apartment locations. A popup callout on the map reads 'Williamsburg / Greenpoint — Studio with backyard or Semi studio — badge-green: Within Budget.' Below the map, a sortable table lists apartment details including name, neighborhood, type, price, and utilities. A legend at the bottom identifies neighborhoods by color: Williamsburg/Greenpoint, Bushwick, Astoria, and table. Right — IA Panel: The Hierarchy tab is active, showing seven 'includes' rules under the 'apartments' entity: Proximity, Furnishing, Internet, Utilities, Type, Price, and Neighborhood, with an '+ Add hierarchy' option. Below, a rule format legend illustrates how each rule type is displayed: Partition (Neighborhood: East village, Upper East), Hierarchy (apartments includes price, includes distance), Order (proximity to work > price), and Vocab ('affordable' defined as under \$1800).
}
    \label{fig:maru-interface}
\end{figure}

\clearpage
\subsection{Layouts and Interactions Supported by \sysname{}}
\label{appendix:supported-interactions}

\begin{table}[h]
\centering
\caption{The 16 layout types supported by \sysname{}. Components can be composed (e.g., map + tabs, tabs containing card-grid).}
\Description{A two-column table listing 16 layout types supported by the system. Each row contains a layout name and its description: table (sortable grid for uniform items), card-grid (browsable grid with detail fields), tabs (grouped content in selectable sections), accordion (collapsible hierarchical sections), kanban (board-style columns for drag-and-drop), timeline (sequential/chronological display), map (spatial layout for location-based data), ranked-list (ordered list by priority), tree-view (hierarchical tree for nested relationships), diagram (node-link graph for connections), calendar (time-based grid for scheduling), hero (large featured display for fewer than 5 items), carousel (sliding card display), masonry (variable-height grid), matrix (multi-dimensional comparison grid), and chart (bar, line, or pie data visualization).
}
\label{tab:layouts}
\small
\begin{tabularx}{\linewidth}{p{0.14\linewidth}X}
\toprule
\textbf{Layout} & \textbf{Description} \\
\midrule
table & Grid layout with sortable columns for uniform items \\
card-grid & Browsable grid of card-style items with detail fields \\
tabs & Tab navigation grouping content into selectable sections \\
accordion & Collapsible sections for hierarchical content exploration \\
kanban & Board-style columns for drag-and-drop organization \\
timeline & Sequential/chronological display of events \\
map & Spatial/geographic layout for location-based data \\
ranked-list & Ordered list emphasizing ranking and priority \\
tree-view & Hierarchical tree for nested relationships \\
diagram & Node-link graph for relationships and connections \\
calendar & Time-based grid view for scheduling \\
hero & Large featured item display (fewer than 5 items) \\
carousel & Sliding/rotating card display \\
masonry & Grid layout with variable height items \\
matrix & Multi-dimensional comparison grid \\
chart & Data visualization (bar, line, pie) \\
\bottomrule
\end{tabularx}
\end{table}

\begin{table}[h]
\centering
\caption{The 10 supported interaction types and their mappings to IA rule dimensions.}
\Description{A three-column table mapping 10 interaction types to IA rule dimensions. Columns are Interaction, Rule Type, and Description. Sort and Move map to Order; field-star maps to Order (field importance). Delete, card-move, and lasso-group map to Partition. Field-remove and field-add map to Hierarchy. Apply-annotation maps to Vocabulary. Annotate is logged only with no rule created until applied.
}
\label{tab:interactions}
\small
\begin{tabularx}{\linewidth}{p{0.18\linewidth}p{0.16\linewidth}X}
\toprule
\textbf{Interaction} & \textbf{Rule Type} & \textbf{Description} \\
\midrule
sort & Order & Sort items by a field. Creates an ordering preference. \\
move & Order & Reorder an item within a group. Creates a relative ranking rule. \\
field-star & Order & Star/prioritize a field. Creates a field importance rule. \\
delete & Partition & Remove item from view. Updates partition membership. \\
card-move & Partition & Move item between groups. Updates source/target partitions. \\
lasso-group & Partition & Create new group via lasso selection. Creates a new partition. \\
field-remove & Hierarchy & Remove a field from an item. Creates a hierarchy-excludes rule. \\
field-add & Hierarchy & Add a new field to items. Creates a hierarchy-includes rule; the LLM fills values. \\
apply-annotation & Vocabulary & Apply a term annotation to selected text. Creates a vocabulary definition. \\
annotate & (logged only) & Open annotation popover. No rule created until applied. \\
\bottomrule
\end{tabularx}
\end{table}

\clearpage

\clearpage
\section{User Study Details}

\subsection{Procedure}
\begin{table}[H]
\centering
\caption{User Study Procedure. Each study session lasted approximately 120 minutes in total.}
\Description{A two-column table showing the user study procedure across seven steps totaling approximately 120 minutes. Steps are: Introduction and interface tutorial (20 min), Task Session S1 (25 min), Debrief with questionnaire and interview (15 min), Task Session S2 (25 min), Debrief with questionnaire and interview (15 min), Task Session S3 (20 min), and Final Interview (20 min).

}
\label{tab:study}
\small
\begin{tabularx}{\linewidth}{p{0.12\linewidth}X}
\toprule
\textbf{Step (min.)} & \textbf{Activity} \\
\midrule
1 \hfill(20) & Introduction on research and tutorial for interface \\
2 \hfill(25) & Task Session (S1) \\
3 \hfill(15) & Debrief Session (questionnaire and brief debrief interview) \\
4 \hfill(25) & Task Session (S2) \\
5 \hfill(15) & Debrief Session (questionnaire and brief debrief interview) \\
6 \hfill(20) & Task Session (S3) \\
7 \hfill(20) & Final Interview \\
\bottomrule
\end{tabularx}
\end{table}

\subsection{Session Task}
The two assigned tasks for first two sessions (S1 and S2) represented common information-intensive scenarios:

\begin{quote}
\textbf{Task 1}
You are considering your next step---applying to graduate programs or exploring job opportunities. Research and compare at least 3–4 options across dimensions like research fit or job role, location, funding or salary, and long-term alignment. By the end of the session, have a shortlist with a clear sense of which options you'd prioritize and why.
\end{quote}

\begin{quote}
\textbf{Task 2}
You are organizing an outdoor picnic for your lab or team of about 15 people. Plan everything out---food and drinks, activities, supplies, a day timeline, and a budget breakdown across people. By the end of the session, have a concrete plan including a shopping/preparation list, a rough activity timeline, and a budget breakdown.
\end{quote}

\subsection{Post-Session Survey}

The post-session survey was administered after each of the two sessions (Session~1 and Session~2) using the same instrument. It consisted of two parts: the NASA Task Load Index (NASA-TLX) and a set of custom items assessing structural alignment.

\begin{table}[H]
\centering
\caption{NASA-TLX Items (administered after each session, 7-point scale: Very Low -- Very High)}
\Description{A two-column table listing the six NASA-TLX workload dimensions and their corresponding survey questions, administered on a 7-point scale from Very Low to Very High. Dimensions are Mental Demand, Physical Demand, Temporal Demand, Performance, Effort, and Frustration.
}
\label{tab:nasa-tlx}
\small
\begin{tabularx}{\linewidth}{p{0.28\linewidth}X}
\toprule
\textbf{Dimension} & \textbf{Item} \\
\midrule
Mental Demand & How much mental and perceptual activity was required? \\
Physical Demand & How much physical activity was required? \\
Temporal Demand & How much time pressure did you feel due to the pace of the task? \\
Performance & How successful were you in accomplishing the task goals? \\
Effort & How hard did you have to work to attain your level of performance? \\
Frustration & How irritated, stressed, and annoyed vs. content were you? \\
\bottomrule
\end{tabularx}
\end{table}

\begin{table}[H]
\centering
\caption{Custom Structural Alignment Items (administered after each session, 5-point scale: 1 -- 5)}
\Description{A two-column table listing five custom structural alignment survey items (C1–C5) administered on a 5-point scale. Items assess whether the output was structured as wanted (C1), whether the user felt in control of organization (C2), whether the output felt tailored (C3), whether the user had to work hard for the desired structure (C4), and whether the system understood the user's needs (C5).
}
\label{tab:custom-session}
\small
\begin{tabularx}{\linewidth}{p{0.10\linewidth}X}
\toprule
\textbf{ID} & \textbf{Item} \\
\midrule
C1 & The output was structured the way I wanted. \\
C2 & I felt in control of how the output was organized. \\
C3 & The output felt tailored to my preferences. \\
C4 & I had to work hard to get the structure I wanted. \\
C5 & The system understood what I needed. \\
\bottomrule
\end{tabularx}
\end{table}

\subsection{Post-Comparison and Final Survey}

After completing both sessions, participants answered a post-comparison question and then engaged in a freeform session with the full system, after which a final survey was administered.

\begin{table}[H]
\centering
\caption{Post-Comparison and Final Survey Items}
\Description{A three-column table of post-comparison and final survey items with columns for Section, Item, and Type. The Post-Comparison section asks which condition felt more aligned (Choice) and why (Open). The Final Survey section asks about the task completed (Open) and three 5-point scale items: whether outputs evolved to match preferences, whether the system learned the user's organizational preferences, and whether the user would use the system for real tasks.
}
\label{tab:final-survey}
\small
\begin{tabularx}{\linewidth}{p{0.22\linewidth}Xp{0.12\linewidth}}
\toprule
\textbf{Section} & \textbf{Item} & \textbf{Type} \\
\midrule
\multirow{2}{*}{Post-Comparison}
  & Which condition felt more aligned with your preferences? & Choice \\
  & Why? & Open \\
\midrule
\multirow{4}{*}{Final Survey}
  & What was the task you did? & Open \\
  & The system's outputs evolved to better match my preferences over time. & 5-point \\
  & I felt the system learned how I wanted information organized. & 5-point \\
  & I would use this system for real information tasks. & 5-point \\
\bottomrule
\end{tabularx}
\end{table}

\subsection{Tasks Selected for S3 (Freeform)}
\begin{table}[h]
\centering
\caption{S3 Self-Selected Tasks}
\Description{A two-column table listing the self-selected freeform tasks for 12 participants in Session 3. Tasks include trip planning to Italy, St. Tropez, and Spain (P1); finding a gift for a partner (P2); shopping for clothes (P3); planning a trip to Hong Kong (P4); a learning plan for beamforming technology (P5); a trip to Barcelona (P6); creating a pescetarian diet plan (P7); planning a fixed weekly schedule for 4 months (P8); trip planning (P9); planning weekend activities in Seoul (P10); a business trip to Barcelona (P11); and a 7-day trip to Barcelona (P12).
}
\label{tab:s3-tasks}
\small
\begin{tabularx}{\linewidth}{p{0.15\linewidth}X}
\toprule
\textbf{Participant} & \textbf{Task} \\
\midrule
P1 & Planning a trip to Italy, St.\ Tropez, and Spain \\
P2 & Finding a gift for a partner \\
P3 & Shopping for clothes \\
P4 & Planning a trip to Hong Kong \\
P5 & Learning plan for beamforming technology \\
P6 & Planning a trip to Barcelona \\
P7 & Creating a pescetarian diet plan \\
P8 & Planning a fixed weekly schedule for 4 months \\
P9 & Trip planning \\
P10 & Planning weekend activities in Seoul \\
P11 & Planning a business trip to Barcelona \\
P12 & Planning a 7-day trip to Barcelona \\
\bottomrule
\end{tabularx}
\end{table}

\clearpage
\subsection{Baseline and System Prompts}

The baseline and \sysname{} conditions shared the same data acquisition, 
rendering pipeline, and LLM model. The baseline passed user interactions as a flat 
timestamped log, \sysname{} encoded the same interactions as typed IA 
rules. Table~\ref{tab:baseline-comparison} summarizes the differences 
across pipeline stages, and the prompt excerpts below illustrate the 
contrast.

Both conditions received identical user 
interactions. The baseline planner prompt included them as:

\begin{quote}
\texttt{[user query]}\\
\texttt{}\\
\texttt{\#\# User's prior interactions}\\
\texttt{[10:32:15] sorted by price}\\
\texttt{[10:33:01] deleted "Hotel C"}\\
\texttt{[10:34:22] moved "Hostel A" to "Budget"}
\end{quote}

\noindent The \sysname{} planner prompt included the same interactions as structured rules:

\begin{quote}
\texttt{The user has configured the following IA preferences:}\\
\texttt{}\\
\texttt{Ordering criteria:}\\
\texttt{~~1. price: User prioritizes cost-effectiveness}\\
\texttt{}\\
\texttt{Partitions:}\\
\texttt{~~- Budget: Hostel A, Hostel B}\\
\texttt{~~- Luxury: Hotel D}\\
\texttt{}\\
\texttt{Hierarchy relationships:}\\
\texttt{~~- "item" -{}-excludes-{}-> "Hotel C"}\\
\texttt{}\\
\texttt{IMPORTANT: Use these consistent field names: price, rating, location}
\end{quote}

\begin{table*}[h]
\centering
\small
\begin{tabular}{p{2.8cm} p{5cm} p{5cm}}
\toprule
\textbf{Stage} & \textbf{Baseline} & \textbf{\sysname{}} \\
\midrule
Rule detection & Skipped & LLM detects IA rules from user query \\
\addlinespace
Rule retrieval & Skipped & LLM retrieves relevant rules by semantic match \\
\addlinespace
Rule scaffolding & Skipped & Detects implied partition/hierarchy from new data \\
\addlinespace
Codebook pipeline & Skipped & Maps user need to display and interaction patterns \\
\addlinespace
Schema hint & None — field names decided freely by LLM & Built from hierarchy rules: consistent field names enforced \\
\addlinespace
IA rules in planner & Empty & Formatted typed rules: orders, vocabularies, partitions, hierarchies \\
\addlinespace
Interaction history & Appended as flat timestamped log & Encoded as typed IA rules \\
\addlinespace
Interaction → rules & Logged as timestamped strings & Creates typed Order/Partition/Hierarchy/Vocabulary rules \\
\addlinespace
Post-generation extraction & Skipped & Extracts hierarchy rules from plan's field structure \\
\bottomrule
\end{tabular}
\caption{Pipeline differences between the baseline and \sysname{} conditions.}
\label{tab:baseline-comparison}
\Description{A three-column table comparing the baseline and Maru pipeline stages across nine dimensions. In the baseline, rule detection, retrieval, scaffolding, codebook pipeline, and post-generation extraction are all skipped; schema hints are absent with field names decided freely by the LLM; IA rules in the planner are empty; interaction history is appended as a flat timestamped log; and interactions are logged as timestamped strings. In Maru, the LLM detects and retrieves IA rules, scaffolds implied partitions and hierarchies, maps user needs via codebook, enforces consistent field names from hierarchy rules, formats typed rules for the planner, encodes interactions as typed IA rules, and extracts hierarchy rules from the generated plan's field structure.
}
\end{table*}